\documentclass[runningheads]{llncs}
\usepackage[T1]{fontenc}
\usepackage{graphicx}
\usepackage{booktabs}
\usepackage[misc]{ifsym}

\usepackage[cmyk]{xcolor}
\usepackage{mwe}
\usepackage{multirow}
\usepackage[hidelinks]{hyperref}
\usepackage{subcaption}
\usepackage{amssymb}

\begin{document}

\title{Collaborative Interest-aware Graph Learning for Group Identification}

\author{Rui~Zhao \inst{1,2} \and
Beihong~Jin \Letter \inst{1,2}  \and
Beibei~Li \inst{3} \and
Yiyuan~Zheng\inst{1,2}}



\institute{Institute of Software, Chinese Academy of Sciences, Beijing, China \and
University of Chinese Academy of Sciences, Beijing, China \and College of Computer Science, Chongqing University, Chongqing, China
\\
\email{Beihong@iscas.ac.cn}}

\authorrunning{R. Zhao et al.}

\titlerunning{Collaborative Interest-aware Graph Learning for Group Identification}

\maketitle              

\begin{abstract}
With the popularity of social media, an increasing number of users are joining group activities on online social platforms. 
This elicits the requirement of group identification (GI), which is to recommend groups to users.
We reveal that users are influenced by both group-level and item-level interests, and these dual-level interests have a collaborative evolution relationship: joining a group expands the user's item interests, further prompting the user to join  new groups. Ultimately, the two interests tend to align dynamically. However, existing GI methods fail to fully model this collaborative evolution relationship, ignoring the enhancement of group-level interests on item-level interests, and suffering from false-negative samples when aligning cross-level interests. In order to fully model the collaborative evolution relationship between dual-level user interests, we propose \textbf{CI4GI}, a \textbf{C}ollaborative \textbf{I}nterest-aware model for \textbf{G}roup \textbf{I}dentification. Specifically, we design an interest enhancement strategy that identifies additional interests of users from the items interacted with by the groups they have joined as a supplement to item-level interests. In addition, 
we adopt the distance between interest distributions of two users to optimize the identification of negative samples for a user,
mitigating the interference of false-negative samples during cross-level interests alignment. The results of experiments on three real-world datasets demonstrate that 
CI4GI significantly outperforms state-of-the-art models.

\end{abstract}

\section{Introduction}
With the proliferation of social media, joining online groups has become a vital way for users to share experiences, explore interests and expand social connections. For example, on the game platform \textit{Steam}, players participate in multiplayer battles by joining game groups. Similarly, on the travel community \textit{Mafengwo}, users can join groups of interest, find travel partners within the groups and plan group trips. For users, group participation serves not only as an effective channel for accessing vertical domain knowledge but also as a critical hub for establishing social bonds. For the platform, users' interests in groups can enhance their engagement and retention. 
Therefore, the group identification (GI) task, i.e., recommending groups to users becomes a topic that needs to be explored.
Compared with directly recommending items to users, recommending groups to users can establish an emotional link between the platform and users and maintain long-term user stickiness.
This paper focuses on the GI task.

We note that the essence of the GI task is to understand users' interests and find groups that attract them. Specifically, users are jointly influenced by two levels of interests when joining groups: group-level interests and item-level interests, which can be learned from their historical participation in groups and historical interactions with items, respectively. Most importantly, these dual-level interests are deeply intertwined and exhibit a collaborative evolution mechanism: users initially select groups based on existing item-level interests, while their group interactions subsequently expand their item-level interests, because the items interacted with by the group might become potential item-level interests for users.
This item-level interest evolution drives users to join more relevant groups,  ultimately leading to a dynamic alignment between the dual-level interests.

\begin{figure}[t]
    \centering
    \includegraphics[width=0.88\linewidth]{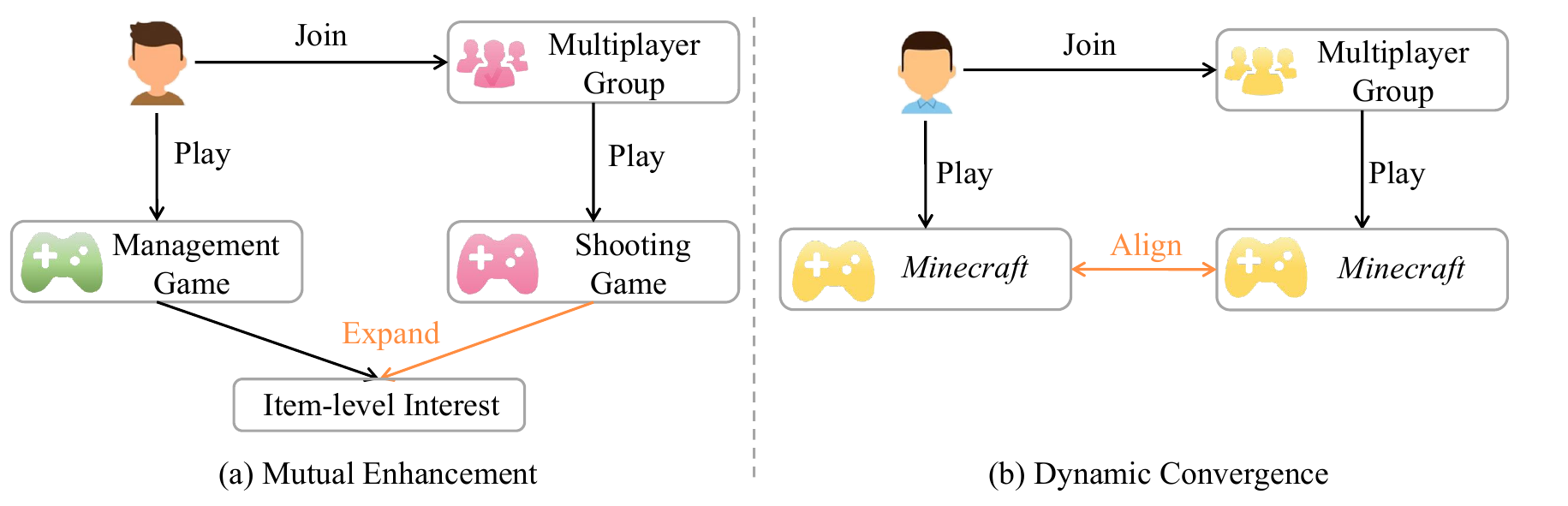}
    \caption{Illustration of mutual enhancement and dynamic convergence.}
    \label{fig:intro}
\end{figure}

The collaborative evolution process between the dual-level interests
plays a crucial role in accurately modeling user interests for the GI task. It can be modeled from two aspects: 
i)\textbf{Mutual Enhancement}: On the one hand, joining groups expands users' item-level interests. For example, in Fig. \ref{fig:intro}(a), a user who typically plays single-player management games also joins a group to participate in multiplayer shooting games. In this case, the user's interest in shooting games within the group becomes a part of the item-level interest. On the other hand, item-level interests also influence users' decisions to join groups. Considering users' item-wise preferences in GI helps uncover groups that users are interested in. ii)\textbf{Dynamic Convergence}: There is an overlap between users' group-level and item-level interests. As shown in Fig. \ref{fig:intro}(b), a user plays \textit{Minecraft} both in single-player mode and within a group in multiplayer mode. At this point, the user's item-level and group-level interests are highly aligned. This property justifies constructing cross-level preference alignment via self-supervised learning, thereby enhancing representation quality and recommendation accuracy.

Existing methods fail to effectively solve the GI task. Traditional recommendation models focus on binary recommendation tasks, such as user-item recommendation, which only model users' item-level interests and are not suitable for group identification. Recently, several methods specifically designed for the GI task have been proposed, such as DiRec~\cite{DiRec} and GTGS~\cite{GTGS}. While these methods model both users’ group-level and item-level interests, they fail to fully capture the collaborative evolution relationship between these dual-level interests. Specifically, they model users’ group-level and item-level interests separately, overlooking the enhancement effect of group participation on users' item-level interests. What is worse, although these methods employ contrastive learning to align cross-level interests within a user and push away this user's interests from other users' interests,
they overlook interest overlaps among similar users, crudely pushing away the representations of similar users, which adversely affects the representation learning and recommendation effectiveness.

To fully model the collaborative evolution relationship between users’ dual-level interests, we propose a collaborative interest-aware graph learning model CI4GI. CI4GI leverages the hypergraph convolution network and graph attention network (GAT) to learn users’ group-level and item-level interests from their historical user-group and user-item interactions. To model the mutual enhancement between these dual-level interests, we design an interest enhancement strategy with two key innovations: i) Item Representation Enhancement: we augment the item representation in the user-item interaction graph based on group-item interaction information. ii) Contextual Enhancement: The group a user joins is considered as the context of that user. We identify additional interests from items interacted with by the groups the user has joined and use them as a supplement to item-level interests. Furthermore, to mitigate false-negative interference in contrastive learning caused by users with similar interests, we propose a dynamic false-negative sample optimized self-supervised learning loss to effectively align users’ group-level and item-level interests.

Our contributions are summarized as follows.

\begin{itemize}
\item[$\bullet$] We highlight that the key to the GI task lies in understanding users' dual-level interests (i.e., group-level and item-level interests) and capturing their collaborative evolution relationship. To address this, we propose a collaborative interest-aware graph learning model CI4GI. 
\item[$\bullet$] We design an interest enhancement strategy that supplements users' item-level interests through item representation enhancement and contextual enhancement.
\item[$\bullet$] We propose a dynamic false-negative sample identification method based on the distance between interest distributions of two users
to alleviate the problem of false-negative samples caused by overlapping interests of similar users.
\item[$\bullet$] We conduct extensive experiments on three publicly available datasets, and the significant improvement of CI4GI on all datasets demonstrates its strength in completing the GI task. 
\end{itemize}

The rest of the paper is organized as follows. Section~\ref{Related Work} introduces the related work and Section~\ref{Methodology} describes the model CI4GI in detail. Subsequently, Section~\ref{Experiments} gives the experimental evaluation. Finally, the paper is concluded in Section~\ref{Conclusion}.

\section{Related Work}
\label{Related Work}

The group identification task is first proposed in CFAG~\cite{CFAG}, and there exist closely related tasks in the name of group recommendation. Therefore, we provide a brief introduction to these tasks and emphasize their differences.


The group recommendation aims to recommend items
for a group of members~\cite{CubeRec,SIGR,HL4EGR}. Existing group recommendation models focus on aggregating the interests of different members to recognize their common interests. Multiple group recommendation models such as AGREE~\cite{AGREE} and SoAGREE~\cite{SoAGREE} propose different attention-based aggregation methods. Recently, ConsRec \cite{ConsRec} models users and items as nodes, groups as hyperedges, and learns group representations through the hypergraph neural network. However, these methods mainly model the item-level interests of users, ignoring users' group-level interests.

In addition to recommending items to groups, the term 'group recommendation' in the literature is also used to denote the recommendation of groups to their potential members. 
Traditional approaches typically use various algorithms to reconstruct user-group membership matrices by utilizing additional auxiliary information, such as the semantic content of group descriptions in CCF~\cite{CCF}, visual information from photos in JTM~\cite{JTM}, and user behaviors across different time periods in DMF~\cite{DMF}.
CFAG~\cite{CFAG} is the first to define the GI task, which learns the interaction relationships among users, groups, and items through tripartite graph convolution layers. DiRec~\cite{DiRec} classifies the user's intention of joining a group into social intention and personal interest intention, and then combines user and group representations under these two intention categories for recommendations. GTGS~\cite{GTGS} models the user-group-item relationships using three hypergraphs: a group hypergraph from the user's perspective, a user hypergraph from the item's perspective, and a user hypergraph from the group's perspective. 
However, all of these approaches model users' group-level and item-level interests separately, not only ignoring the extension of group-level interests to item-level interests but also often improperly aligning two interests through contrastive learning.

On the other hand, our work is related to graph-based recommender systems (RSs), which can be broadly categorized into three classes from a modeling perspective: (i) graph convolutional network-based RSs~\cite{LightGCN,NGCF,SGL}; (ii) graph attention network-based RSs~\cite{GAT,MGAT}; and (iii) gated graph neural network-based RSs~\cite{g-3}. Most graph-based RSs focus only on user-item bipartite graphs, and directly applying these models to the group identification task is inappropriate due to the fact that they cannot capture the complex relationships among users, items, and groups. To realize the GI task, new graph-based models need to be developed.

\section{Methodology}
\label{Methodology}

\subsection{Model Overview}
Let $\mathcal{U}$, $\mathcal{V}$ and $\mathcal{G}$ denote the user set, item set, and group set, respectively. There are three types of observed interactions among users, items, and groups, i.e., user-item interactions denoted as $\mathbf{X} \in \mathbb{R}^{|\mathcal{U}|\times|\mathcal{V}|}$, group-item interactions represented as $\mathbf{Y} \in \mathbb{R}^{|\mathcal{G}|\times|\mathcal{V}|}$, and user-group affiliations represented as $\mathbf{Z} \in \mathbb{R}^{|\mathcal{U}|\times|\mathcal{G}|}$. Given a user $u_i \in \mathcal{U}$, the goal of the group identification task is to predict the groups that user $u_i$ has not yet joined but is highly likely to be interested in joining in the future.

\begin{figure}[t]
    \centering
    \includegraphics[width=1\linewidth]{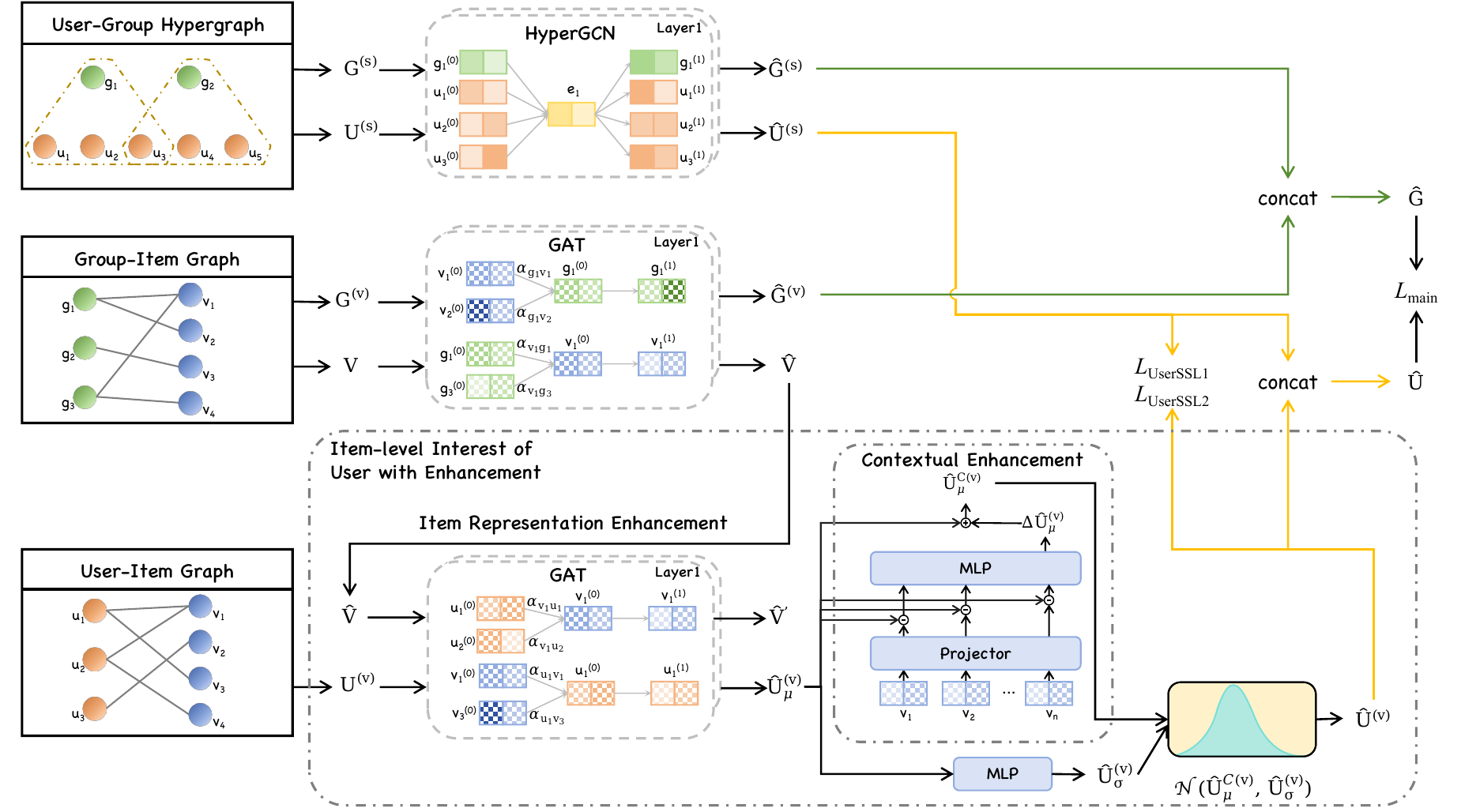}
    \caption{Architecture of CI4GI.}
    \label{fig:model}
\end{figure}

For this task, we propose a collaborative interest-aware graph learning model CI4GI, 
whose architecture is shown in Fig. \ref{fig:model}. CI4GI first employs a hypergraph to model user-group history interactions to learn group-level interests for users and groups. 
Then, CI4GI leverages two graph attention networks to model user-item and group-item interactions, capturing item-level interests for users and groups.
To model the mutual enhancement between these dual-level interests,
we design an interest enhancement strategy with two types of enhancements. First, item representation enhancement 
leverages group-item interaction information to augment item representations that have learned via the user-item interaction graphs.  
Second, contextual enhancement identifies additional interests of users from the items interacted with by the groups they have joined, serving as a supplement to users' item-level interests. Further, we propose a contrastive learning loss with dynamic false-negative sample optimization. It aligns users' group-level and item-level interests while alleviating the problem of false-negative samples caused by overlapping interests of similar users.

\subsection{Embedding Layer}
We maintain an embedding layer to initialize the learnable embeddings of users, items, and groups. For item set $\mathcal{V}$, the learnable item embedding matrix $\mathbf{V} \in \mathbb{R}^{|\mathcal{V}|\times d}$ is obtained after the embedding layer, where $d$ is the embedding size. For the users, we represent group-level and item-level interests in two matrices, i.e., the group-level interest matrix $\mathbf{U}^{(s)} \in \mathbb{R}^{|\mathcal{U}|\times d}$ and the item-level interest matrix $\mathbf{U}^{(v)} \in \mathbb {R}^{|\mathcal{U}|\times d}$. Similarly, for groups, there are $\mathbf{G}^{(s)} \in \mathbb{R}^{|\mathcal{G}|\times d}$ and $\mathbf{G}^{(v)} \in \mathbb{R}^{|\mathcal{G}|\times d}$.

\subsection{Group-level Interest Learning}
We employ a hypergraph to model user-group affiliations. 
CI4GI defines a user-group affiliation hypergraph as $H^{UG} = (\mathcal{N},\mathcal{E}^{UG})$, where $\mathcal{N} = \mathcal{U} \cup \mathcal{G}$ denotes the node set, i.e., a node of $H^{UG}$ is a user in $\mathcal{U}$ or a group in $\mathcal{G}$. 
The hyperedge set is denoted as $\mathcal{E}^{UG}$, where a hyperedge $e_k^{UG} \in \mathcal{E}^{UG}$, $k \in[1, |\mathcal{G}|]$ connects all user nodes in the group $g_k$ along with the group node $g_k$.
The association matrix $T^{UG} = [t_{ik}^{UG}] \in \mathbb{R}^{(|\mathcal{U}| + |\mathcal{G}|) \times |\mathcal{G}|}$ represents the connectivity between nodes and hyperedges in the hypergraph $H^{UG}$.

By modeling the hypergraph with a hyperedge connecting the group and all its members, we preserve the social relationships within the group. 
We then employ the classical hypergraph convolution for representation learning:
\begin{equation}
    N^{(l)} = \sigma (D_{\mathcal{N}}^{-\frac{1}{2}} T^{UG} D_{\mathcal{E}}^{-1} {T^{UG}}^{T} D_{\mathcal{N}}^{-\frac{1}{2}}N^{(l-1)} \mathrm{W}^{UG})
\end{equation}
where $D_{\mathcal{N}} \in \mathbb{R}^{|\mathcal{N}|\times|\mathcal{N}|}$ and $D_{\mathcal{E}}^{|\mathcal{E}|\times|\mathcal{E}|}$ represent the degree matrices of the nodes and hyperedges, respectively. 
$N^{(l)} \in \mathbb{R}^{|\mathcal{N}|\times d}$ is the node representation of the $l$-th layer, $N^{(0)}=\mathrm{Concat}(\mathbf{U}^{(s)},\mathbf{G}^{(s)})$, $\mathrm{W}^{UG} \in\mathbb{R}^{d \times d}$ is a learnable matrix, $\sigma$ is the sigmoid function.

Finally, the group-level interests of users and groups are obtained after $L$ layers hypergraph convolution:
$\hat{\mathbf{U}}^{(s)}, \hat{\mathbf{G}}^{(s)} = N^{(L)}$.

\subsection{Item-level Interest Learning}
We learn the item-level interests of groups and users, respectively.

\noindent \textbf{Item-level Interest of Group.}
We define the group-item interaction graph as $G^{GV} = (\mathcal{N}^{GV}, \mathcal{E}^{GV})$, where $\mathcal{N}^{GV} = \mathcal{G} \cup \mathcal{V}$ denotes the node set, an edge $e_{jk}^{GV} \in \mathcal{E}^ {GV}, j \in [1,|\mathcal{V}|], k \in [1,|\mathcal{G}|]$ denotes the group $g_k$ interacting the items $v_j$. The association matrix $T^{GV} \in \mathbb{R}^{(|\mathcal{G}|+|\mathcal{V}|) \times (|\mathcal{G}|+|\mathcal{V}|)}$ represents the connectivity between nodes in $G^{GV}$. We use classical GAT~\cite{GAT} for representation learning:
\begin{equation}
    \hat{\mathbf{G}}^{(v)}, \hat{\mathbf{V}} = \mathrm{GAT}_{\Theta_{GV}}([\mathbf{G}^{(v)}||\mathbf{V}], T^{GV})
\end{equation}
where $\mathrm{GAT}_{\Theta_{GV}}$ is a stack of $L$-layer graph attention networks, $\mathbf{G}^{(v)}$ and $\mathbf{V}$ are the initial
groups' item-level interest representations and item representations, $\hat{\mathbf{G}}^{(v)} \in \mathbb {R}^{|\mathcal{G}| \times d}$ and $ \hat{\mathbf{V}} \in \mathbb{R}^{|\mathcal{V}| \times d}$ are those learned from GAT.

\noindent \textbf{Item-level Interest of User with Enhancement.}
Similarly, CI4GI defines the user-item interaction graph as $G^{UV} = (\mathcal{N}^{UV}, \mathcal{E}^{UV})$. The association matrix $T^{UV} \in \mathbb{R}^{(|\mathcal{U}|+|\mathcal{V}|) \times (|\mathcal{U}|+|\mathcal{V}|)}$ denotes the connectivity between nodes in $G^{UV}$. We also apply a graph attention network to learn user-item interactions.

Further, we propose an interest enhancement strategy with two methods: item representation enhancement and contextual enhancement.

\textit{\textbf{Item representation enhancement}} aims to enhance the item representation in user-item interaction graph with the group-item interaction information,
thereby capturing the similarity between the item-level interest of the group and the user.
This helps the model recognize the user's tendency to join a group based on their interests in the items. Specifically, we use the item representation $\hat{\mathbf{V}}$ learned from $G^{GV}$ as the initial item representation for $G^{UV}$:
\begin{equation}
\label{mu}
    \hat{\mathbf{U}}^{(v)}_\mu, \hat{\mathbf{V'}} = \mathrm{GAT}_{{\Theta_{UV}}}([\mathbf{U}^{(v)}||\hat{\mathbf{V}}], T^{UV})
\end{equation}
where $\mathrm{GAT}_{{\Theta_{UV}}}$ is the graph attention network, $\mathbf{U}^{(v)}$ is the initial users' item-level interest representations and $\hat{\mathbf{V}}$ is the item representation learned from $G^{GV}$. $\hat{\mathbf{U}}^{(v)}_\mu$ and $\hat{\mathbf{V'}}$ are the user and item representations learned by the GAT.

In group identification scenarios, users' item-level interests are extended by joining groups, with items shared within these groups becoming potential interests for the user. 
Therefore, we propose a \textit{\textbf{contextual enhancement}} that treats the groups the user has joined as their context information, identifying potential interests from the items that the user has historically interacted with by joining the group.
Specifically, we first identify contextual items by the user-group affiliation matrix $\mathbf{Z}$ and the group-item interaction matrix $\mathbf{Y}$. Then, we project the contextual item representation $\hat{\mathbf{V}}$ into the same vector space as the user's representation $\hat{\mathbf{U}}^{(v)}_\mu$ using a projection matrix $\mathrm{W}^C$. Finally, we compute the distance between the user representation and the contextual items embeddings to obtain the increment $\Delta \hat{\mathbf{U}}^{(v)}_\mu$ with the following formula:
\begin{equation}
    \Delta \hat{\mathbf{U}}^{(v)}_\mu = \mathrm{MLP}(\hat{\mathbf{U}}^{(v)}_\mu - D^{-1}_{ZY} \mathbf{ZY} \times \hat{\mathbf{V}} \mathrm{W}^C )
\end{equation}
where $D_{ZY} \in \mathbb{R}^{|\mathcal{U}| \times |\mathcal{U}|}$ is the row degree matrix of the matrix obtained after multiplying matrices $\mathbf{Z}$ and $\mathbf{Y}$, which is used for normalization. $\mathrm{W}^C \in \mathbb{R}^{d \times d}$ is the projection matrix.

The context-enhanced user representation is then obtained by adding the increment $\Delta \hat{\mathbf{U}}^{(v)}_\mu$:
\begin{equation}
    \hat{\mathbf{U}}^{C(v)}_\mu = \hat{\mathbf{U}}^{(v)}_\mu + \gamma \Delta \hat{\mathbf{U}}^{(v)}_\mu
\end{equation}
where $\gamma$ is a hyperparameter indicating the weight of the increment.

Further, we represent the users' item-level interests as independent Gaussian distributions instead of fixed embeddings, which improves the robustness and flexibility of users' interest representations. 
Specifically, based on the user representation $\hat{\mathbf{U}}^{(v)}_\mu$ learned from Eq. \ref{mu} (which is regarded as the mean of the user interest distribution), the variance of the user interest distribution is computed by a multilayer perceptron: $\hat{\mathbf{U}}^{(v)}_\sigma = \mathrm{MLP}(\hat{\mathbf{U}}^{(v)}_\mu)$.

We sample user representations $\hat{\mathbf{U}}^{(v)}$ from independent Gaussian distributions $\hat{\mathbf{U}}^{(v)} \sim \mathcal{N}(\hat{\mathbf{U}}^{(v)}_\mu, \hat{\mathbf{U}}^{(v)}_\sigma)$. 
Since direct sampling prevents gradient backpropagation, we apply the reparameterization trick as follows:
\begin{equation}
\hat{\mathbf{U}}^{(v)} = \hat{\mathbf{U}}^{C(v)}_\mu + \hat{\mathbf{U}}^{(v)}_\sigma \epsilon
\end{equation}
where $\epsilon$ is randomly sampled from the standard normal distribution $\mathcal{N}(0,1)$.

\subsection{Cross-level Interest Alignment}


In this paper, we model users' group-level and item-level interests separately. Since these interests may overlap in real group identification scenarios, we propose using contrastive learning to align the cross-level interests.
Additionally, similar users' interests can overlap, and vanilla contrastive learning may lead to false-negative samples. To address this, we introduce a dynamic false-negative sample optimized contrastive learning loss, using Wasserstein distance between user interest distributions to identify false-negative samples and improve the quality of cross-level interest alignment.

\noindent\textbf{Vanilla Contrastive Learning.} We align the cross-level interests of the same user by performing conventional contrastive learning on the user's group-level interest $\hat{\mathbf{U}}^{(s)}$ and item-level interest $\hat{\mathbf{U}}^{(v)}$ via the following InfoNCE~\cite{infonce} loss: 
\begin{equation}
\label{eq28}
    L_{\mathrm{UserSSL1}} = -\sum_{u_i \in \mathcal{U}} \log{\frac{\exp(sim(\hat{\mathbf{u}}_{i}^{(v)},\hat{\mathbf{u}}_{i}^{(s)})/{\tau})}{{\exp(sim(\hat{\mathbf{u}}_{i}^{(v)},\hat{\mathbf{u}}_{i}^{(s)})/{\tau})}+N_{\mathrm{V1}}^U+N_{\mathrm{S1}}^U}}
\end{equation}
\begin{equation}
\label{eq29}
    N_{\mathrm{V1}}^U =\!\!\!\sum_{u_{i^{\prime}}\in \mathcal{U}_i^-}\!\!\!{\exp(sim(\hat{\mathbf{u}}_{i^{\prime}}^{(v)},\hat{\mathbf{u}}_{i}^{(s)})/{\tau})},~N_{\mathrm{S1}}^U =\!\!\! \sum_{u_{i^{\prime}}\in \mathcal{U}_i^-}\!\!\!{\exp(sim(\hat{\mathbf{u}}_{i}^{(v)},\hat{\mathbf{u}}_{i^{\prime}}^{(s)})/{\tau})}
\end{equation}
where $\hat{\mathbf{u}}_{i}^{(s)}$ and $\hat{\mathbf{u}}_{i}^{(v)}$ form a pair of positive samples. $\mathcal{U}_i^-$ is the set of negative samples w.r.t. the user $u_i$, which is composed of other users (i.e., $u_{i'}\neq u_i$ ) within the same batch. $sim(\cdot)$ is to calculate the similarity of a pair of vectors, which refers to the cosine similarity in this paper. $\tau$ is the temperature parameter.

\noindent\textbf{Dynamic False-negative Sample Optimization.}
We dynamically identify false-negative samples by comparing users' item-level interest distributions. 
Specifically, for the item-level interest distribution $\hat{\mathbf{u}}_{i}^{(v)}$ of user $u_i$, we compute the Wasserstein distance to the distribution $\hat{\mathbf{u}}_j^{(v)}$ of any other user $u_j$. If the distance is less than or equal to a threshold $\mu$, $u_j$ is considered a false-negative sample of $u_i$. Only users with a Wasserstein distance greater than $\mu$ from $u_i$ are included in the negative sample set.
\begin{equation}
\label{eq32}
    \mathcal{N}eg_i = \{ u_j |~ u_j \in \mathcal{U}~\mathrm{and}~ \mathrm{d_{W2}}(u_i,u_j) > \mu \}
\end{equation}
where $\mathrm{d_{W2}}$ denotes the Wasserstein distance, the higher the similarity of two users' distributions, the smaller the Wasserstein distance, $\mu$ is the threshold, and $\mathcal{N}eg_i$ is the set of negative samples w.r.t. the user $u_i$.

To ensure that the false-negative sample set remains adaptive, we dynamically update the set in each batch based on the latest interest distributions. This allows the contrastive learning process to continuously adjust to evolving user interests and prevents the model from being misled by static or outdated negative samples.
Then, similar to conventional user contrastive learning, the following loss is used to align cross-level interest representations of the same user:
\begin{equation}
\label{eq33}
    L_{\mathrm{UserSSL2}} = -\sum_{u_i \in \mathcal{U}} \log{\frac{\exp(sim(\hat{\mathbf{u}}_{i}^{(v)},\hat{\mathbf{u}}_{i}^{(s)})/{\tau})}{{\exp(sim(\hat{\mathbf{u}}_{i}^{(v)},\hat{\mathbf{u}}_{i}^{(s)})/{\tau})}+N_{\mathrm{V2}}^U+N_{\mathrm{S2}}^U}}
\end{equation}
\begin{equation}
\label{eq34}
    N_{\mathrm{V2}}^U \!=\!\!\!\!\!\!\!\!\! \sum_{u_{i^{\prime}}\in \mathcal{U}_i^- \cap \mathcal{N}eg_i}\!\!\!\!\!\!\!\!\!\!\!{\exp(sim(\hat{\mathbf{u}}_{i^{\prime}}^{(v)},\hat{\mathbf{u}}_{i}^{(s)})/{\tau})}, ~N_{\mathrm{S2}}^U = \!\!\!\!\!\!\!\sum_{u_{i^{\prime}}\in \mathcal{U}_i^- \cap \mathcal{N}eg_i}\!\!\!\!\!\!\!\!\!\!{\exp(sim(\hat{\mathbf{u}}_{i}^{(v)},\hat{\mathbf{u}}_{i^{\prime}}^{(s)})/{\tau})}
\end{equation}

\subsection{Model Optimization}

After obtaining the group-level and item-level interest representations of users and groups, we can calculate the main loss of CI4GI by three steps.
First, the two types of interests are concatenated for both users and groups to obtain the final user representation $\hat{\mathbf{U}}$ and group representation $\hat{\mathbf{G}}$:
\begin{equation}
    \label{eq35}
    \hat{\mathbf{U}} = [\hat{\mathbf{U}}^{(s)}||\hat{\mathbf{U}}^{(v)}],~~~
    \hat{\mathbf{G}} = [\hat{\mathbf{G}}^{(s)}||\hat{\mathbf{G}}^{(v)}]
\end{equation}

Then, the dot product similarity is used to compute the probability score $s_{ik}$ of a user $u_i$ joining the group $g_k$:
\begin{equation}
    \label{eq36}
    s_{ik} = \hat{\mathbf{u}}_i \cdot \hat{\mathbf{g}}_k
\end{equation}
where $\hat{\mathbf{u}}_i = \hat{\mathbf{U}}(i,:)$ is the final representation of the user $u_i$ and $\hat{\mathbf{g}}_k = \hat{\mathbf{G}}(k,:)$ is the final representation of the group $g_k$. Subsequently, we calculate the BPR loss as the main loss:
\begin{equation}
\label{eq37}
    L_{\mathrm{main}} = \frac{1}{|\mathcal{T}|}\sum_{(u_i,g_k,g_{k'})\in \mathcal{T}} -\mathrm{log}~\sigma (s_{ik}-s_{ik'})
\end{equation}
where $\mathcal{T} = \{(u_i,g_k,g_{k'})~|~z_{ik}=1 ~\mathrm{and}~ z_{ik'}=0 \} $ is the training set, $z \in \mathbf{Z}$ is the element in the user-group affiliation matrix. $\sigma$ is the sigmoid function.

Finally, we jointly optimize the main loss and auxiliary losses:
\begin{equation}
\label{eq39}
    L = L_{\mathrm{main}} + \lambda_1(\beta L_{\mathrm{UserSSL1}}  + (1-\beta) L_{\mathrm{UserSSL2}}) + \lambda_2 ||\Theta||_2^2
\end{equation}
where $\Theta$ is all trainable parameters in CI4GI, $||\Theta||_2^2$ is the regularization loss, $\lambda_1$ and $\lambda_2$ are hyperparameters, and $\beta = 1/({1+\mathrm{exp}(-k(epoch - E))}) $ is the annealing parameter that controls the weight of the two types of contrastive learning loss, where $k$ and $E$ are hyperparameters.

\subsection{Complexity Analysis}
\noindent\textbf{Space Complexity.} In CI4GI, the learnable parameters mainly come from item embeddings, two user interest embeddings and two group interest embeddings. In addition, in the hypergraph convolution, the number of parameters of the $L$ layer is $Ld^2$, and the number of parameters of the two $L$ layers GAT is $2L(d^2+2d)$. The number of parameters in the MLP and the projection matrix is $O(d^2+d)$. Therefore, the space complexity of CI4GI is $O(|\mathcal{V}|d+|\mathcal{U}|d+|\mathcal{G}|d+Ld^2+Ld)$.

\noindent\textbf{Time Complexity.} The computational amount during the training of CI4GI is mainly concentrated on the hypergraph convolution and GAT. Assuming that $|T^{UG}|$ denotes the number of non-zero elements of the association matrix $T^{UG}$, the time complexity of hypergraph convolution is $O(L(|\mathcal{U}|+|\mathcal{G}|)d^2+L|T^{UG}|d)$. Assuming that $|T^{GV}|$ and $|T^{UV}|$ denote the number of non-zero elements in the association matrices $T^{GV}$ and $T^{UV}$, respectively, the time complexity of the two GATs is $O(L(|\mathcal{U}|+|\mathcal{G}|+|\mathcal{V}|)d^2+L|(T^{GV }|+|T^{UV}|)d)$. 
Therefore, the time complexity of CI4GI is $O(Ld^2(|\mathcal{U}|+|\mathcal{V}|+|\mathcal{G}|)+Ld(|T^{UG}|+|T^{GV}|+|T^{UV}|))$. 

\section{Experiments}
\label{Experiments}

\begin{table} [t]
\setlength{\tabcolsep}{0.9mm}
    \caption{Statistics of datasets.}
    \label{table1}
    \resizebox{\textwidth}{!}{
    \begin{tabular}{ccccccc}
    \toprule
    Dataset & \# Users & \# Items & \# Groups & \# User-Group & \# User-Item & \# Group-Item  \\
    & & & & \ participation & interactions & interactions \\
    \midrule
    Mafengwo & 1,269 & 999 & 972 &5,574 &8,676 & 2,540\\
    Weeplaces &1,501 &6,406 &4,651 &12,258 &43,942 &6,033 \\
    Douban &11,099 &2,351 &1,085 &57,654& 444,776 &23,318\\ 
    \bottomrule
    \end{tabular}
    }
\end{table}

\subsection{Experimental Setup}

\noindent\textbf{Datasets.} We choose three public datasets to conduct experiments.
\begin{itemize}
\item[$\bullet$] \textbf{Mafengwo.} It records the travel history of users on the Mafengwo APP, where users can create or join groups, taking offline group trips. We use the dataset published by CFAG~\cite{CFAG}.

\item[$\bullet$] \textbf{Weeplaces.}  It records users' check-ins on location-based social networks in major cities of the U.S. We follow the same operations as in GroupIM~\cite{GroupIM} for constructing user-POI interactions and group-POI interactions.

\item[$\bullet$] \textbf{Steam.} It records users' game preferences on the online gaming platform Steam, where users have their records of games they have played and can create or join a group. 
We use the dataset published by CFAG~\cite{CFAG}.
\end{itemize}

Table \ref{table1} lists the statistics of the three datasets. We randomly split the user participation in each dataset into training, validation, and test sets with a ratio of 7:1:2. 

\noindent\textbf{Baselines.} 
The following baselines are chosen to compare with CI4GI:

Three recommendation models: \textbf{LightGCN}\footnote{https://github.com/kuandeng/LightGCN}, which is a classical graph-based recommendation model~\cite{LightGCN}; \textbf{SGL}\footnote{https://github.com/wujcan/SGL-Torch}, which introduces contrastive learning into GNN-based recommendation by generating contrast views through node dropout, edge dropout, or random walk~\cite{SGL}; \textbf{SimGCL}\footnote{https://github.com/Coder-Yu/QRec}, which presents an embedding-based enhancement method to construct positive sample pairs in contrastive learning by adding uniform noise to the embedding~\cite{SimGCL}. As DiRec\cite{DiRec} does, we apply them on the GI task by treating each group as an item and thus only utilizing user-group affiliations. 

Two group recommendation models: \textbf{AGREE}\footnote{https://github.com/LianHaiMiao/Attentive-Group-Recommendation}, which is a classical group recommendation model using an attention mechanism for member aggregation~\cite{AGREE};
\textbf{ConsRec}\footnote{https://github.com/FDUDSDE/WWW2023ConsRec}, the state-of-the-art model for group recommendation, which proposes a hypergraph neural network to learn member-level aggregation and captures the group consensus on three views~\cite{ConsRec}. As DiRec\cite{DiRec} does, we adapt them by replacing their initial group-item BPR loss with user-group BPR loss.

Three GI models: \textbf{CFAG}\footnote{https://github.com/mdyfrank/CFAG}, a classical GI model, which constructs a group-user-item tripartite graph and designs a tripartite graph convolution layer~\cite{CFAG}; 
\textbf{DiRec}\footnote{https://github.com/WxxShirley/CIKM2023DiRec}, which divides the user's intention of joining a group into social intention and personal interest intention~\cite{DiRec}; 
\textbf{GTGS}\footnote{https://github.com/mdyfrank/GTGS}, which 
models the relationships between users, groups, and items by three hypergraphs, and proposes transition hypergraph convolution by using users' preferences for items as a prior knowledge~\cite{GTGS}.

\noindent\textbf{Implementation Details.} We implement our model in PyTorch. In our model, $L$ is set to 2, the temperature $\tau$ is set to 1, the hyperparameter $k$ in $\beta$ is set to 0.1, and the threshold $E$ in $\beta$ is set as follows: 20 for Mafengwo, 30 for Weeplaces and Steam, and contextual enhancement representation weight $\gamma$ is set as follows: 1 for Mafengwo and Steam, and 1.25 for Weeplaces, and the Wasserstein distance threshold $\mu$ is set as follows: 1.5 for Mafengwo, 2.0 for Weeplaces and Steam, $\lambda_1$ and $\lambda_2$ are set to $ 1\times e^{-4}$. For the sake of fairness, we set the size of all embeddings $d$ to 256, the batch size to 1024 and the learning rate to 0.005 in all the experiments. For all baselines, the hyperparameters are set to values corresponding to best performance reported in their respective papers. Experiments are conducted on NVIDIA RTX3090 GPU with 24G memory. The implementation code has been released\footnote{https://github.com/ZhaoRui-7/CI4GI}. 

\noindent\textbf{Metrics.} 
To evaluate the performance of recommending groups to users, we adopt two metrics, i.e., Recall@K and NDCG@K (R@K and N@K for short), where Recall focuses on whether the user actually chooses the recommended group, NDCG focuses on the ranking of the recommended groups and $K$ is set to either 10 or 20.

\subsection{Performance Comparison}
\noindent\textbf{Overall Performance.} Table \ref{table2} lists the experimental results on the three datasets. From Table \ref{table2}, we have the following observations.

\begin{table}[t]
\setlength{\tabcolsep}{0.6mm}
    \centering
    \caption{Overall performance. The values in bold and underlined are the best and second best results in each column.}
     \resizebox{\textwidth}{!}{
    \begin{tabular}{c|cccc|cccc|cccc}
    \toprule
         Dataset & \multicolumn{4}{c|}{Mafengwo} & \multicolumn{4}{c|}{Weeplaces} & \multicolumn{4}{c}{Steam} \\
    \midrule
         Metric & R@10 & R@20 & N@10 & N@20 & R@10 & R@20 & N@10 & N@20 & R@10 & R@20 & N@10 & N@20 \\
    \midrule
        LightGCN &0.2925 &0.3607 &0.1865 &0.2040 &0.2490 &0.3159 &0.1465 &0.1646 &0.2411 &0.3283 &0.1333 &0.1558\\
        SGL &0.2957 &0.3628 &0.1937 &0.2109 &0.2511 &0.3124  &0.1456 &0.1624 &0.2327 &0.3241 &0.1289 &0.1524\\
        SimGCL &0.2943 &0.3576 &0.1890 &0.2052 &0.2486 &0.3140 &0.1466 &0.1645&0.1872 &0.2896 &0.0930 &0.1189\\
    \midrule
        AGREE &0.1679 &0.2302 &0.1061 &0.1222 &0.2312 &0.2957 &0.1355 &0.1529 &0.1882 &0.2977 &0.0996 &0.1275\\
        ConsRec &0.3403 &0.4312 &0.2161 &0.2382 &0.3451 &0.4379 &0.2039 &0.2288 &0.2568 &0.3608 &0.1359 &0.1626\\
    \midrule
        CFAG &0.3007 &0.4051 &0.1698 &0.1965 &0.3848 &0.4778 &0.2251 &0.2529 &0.2328 &0.3427 &0.1224 &0.1507\\
        DiRec &0.3588 &0.4636 &0.2231 &0.2500 &\underline{0.3904} &\underline{0.4843} &\underline{0.2341} &\underline{0.2591} &\underline{0.2738} &\underline{0.3745}  &\underline{0.1436} & \underline{0.1696}\\
        GTGS &\underline{0.3611} &\underline{0.4672} &\underline{0.2248} &\underline{0.2520} &0.3813 &0.4693 &0.2308 &0.2547 &0.2225 &0.3275 &0.1110 &0.1379\\
    \midrule
        CI4GI &\textbf{0.3937} &\textbf{0.5016} &\textbf{0.2557} &\textbf{0.2829} &\textbf{0.4179} &\textbf{0.4990} &\textbf{0.2664} &\textbf{0.2879} &\textbf{0.2912} &\textbf{0.3823} &\textbf{0.1605} &\textbf{0.1840}\\
        Inprov.(\%) &9.02 &7.36 &13.74 &12.26 &7.04 &3.03 &13.79 &11.11 &6.35 &2.08 &11.76 &8.49\\
    \bottomrule
         
    \end{tabular}
    }
    \label{table2}
\end{table}

Traditional recommendation models perform poorly because they typically represent user-group affiliations as graphs, capturing only users' group-level interests while overlooking their item-level interests. Similarly, group recommendation models are not well-suited for the GI task, as they lack the capability to model the complex user interests in GI scenarios.

In contrast, GI models achieve better performance as they are specifically designed for this task. However, while they capture both users' group-level and item-level interests, they overlook the collaborative relationship between the dual-level interests, limiting their effectiveness.

Our CI4GI outperforms all baselines on three datasets. Taking NDCG@10 as an example, compared to the best baseline on each of the three datasets, CI4GI shows improvements of 11.76\% - 13.79\%, averaging at 13.09\%. 


\noindent\textbf{Cold-start Performance.} We evaluate the cold-start performance of CI4GI on Mafengwo and Weeplaces. Specifically, we randomly remove user-group interaction history in the training set, ensuring that each user has joined at most $k$ groups, where $k=1,2,3,4$.
We choose DiRec, GTGS, ConsRec and SGL, the best models from different types of baselines, for comparison.

\begin{figure}[t]
    \centering
    \includegraphics[width=\linewidth]{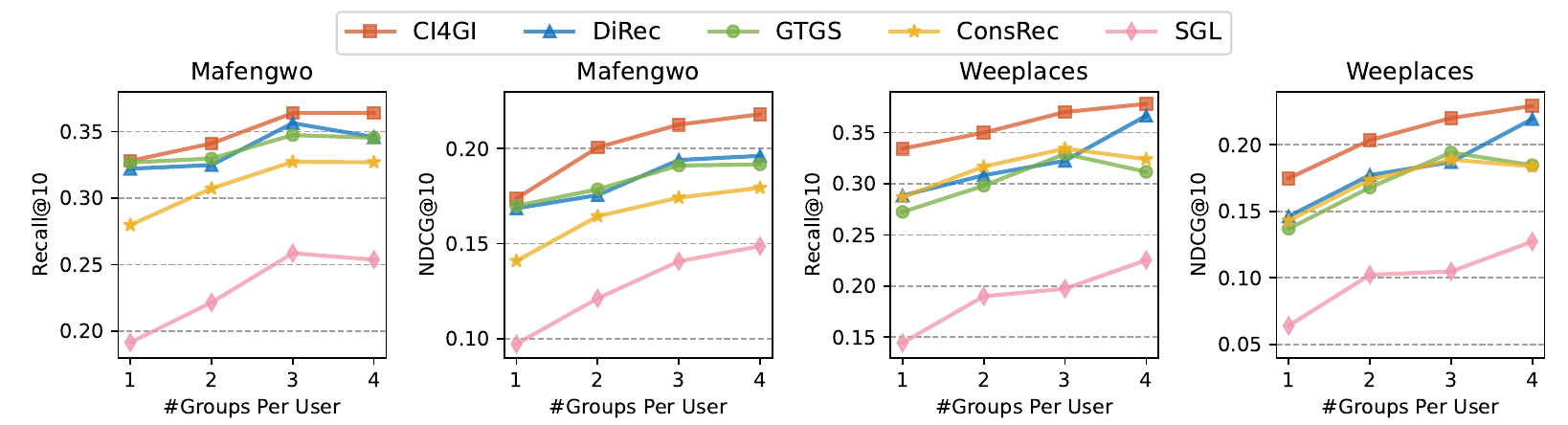}
    \caption{Cold-start performance comparsion.}
    \label{fig:cold-start}
\end{figure}

As shown in Fig. \ref{fig:cold-start}, CI4GI outperforms the baselines in all cases. 
In particular, CI4GI achieves the best performance even when $k=1$, i.e., the user has only one interacted group in history. 
This could be because CI4GI applies an interest enhancement strategy in the user's item-level interest learning, which improves the representation of the user's item-level interest and alleviates the deficiency of the user's group-level interests in cold-start scenarios.
As the threshold $k$
increases from 1 to 3, the performance of all models improves, indicating that more historical user-group interactions are beneficial to interest learning. However, when $k$ increases from 3 to 4, 
some models experience a decline in performance (e.g., DiRec on Mafengwo, GTGS and ConsRec on Weeplaces), while CI4GI continues to improve its performance.
This may be because a higher number of user-group interactions introduce more complex user's interests, and 
CI4GI can capture these intricate interests by identifying the collaborative relationships between dual-level user interests, thus leading to superior performance.


\subsection{Ablation Study}
\begin{table}[t]
    \centering
    \setlength{\tabcolsep}{0.8mm}
    \caption{Ablation study.}
    \label{tab:ablation}
    \begin{tabular}{lcccccc}
         \toprule
          \multirow{2}{*}{Model}& \multicolumn{2}{c}{Mafengwo} & \multicolumn{2}{c}{Weeplaces} & \multicolumn{2}{c}{Steam}\\
        \cmidrule{2-7}
         &R@10 & N@10  & R@10  & N@10 &R@10 & N@10\\
        \midrule
        CI4GI & \textbf{0.3973} & \textbf{0.2557} & \textbf{0.4179} & \textbf{0.2664} & \textbf{0.2912} & \textbf{0.1605} \\
        \midrule
        (A)w/o group-level interests & 0.3351 & 0.2135 & 0.3828 & 0.2349& 0.2699 & 0.1443\\
        (B)w/o item-level interests & 0.3203 & 0.2074 & 0.3097 & 0.1954 & 0.2836 & 0.1529 \\
        \midrule
        (C)w/o enhancement & 0.3687 & 0.2365 & 0.3873 & 0.2436 & 0.2820 & 0.1592 \\
        (D)w/o item enhancement&0.3817 & 0.2547 & 0.4069 & 0.2508 & 0.2866 & 0.1602 \\
        (E)w/o contextual enhancement &0.3697&0.2426 & 0.3894 & 0.2460 & 0.2838 &0.1541 \\
        \midrule
        (F)w/o CL &0.3730 & 0.2484 & 0.3994 &0.2549 &0.2765 &0.1547 \\
        (G)w/o $L_{\mathrm{UserSSL1}}$ &0.3818 & 0.2503 & 0.4134 & 0.2649 & 0.2837 & 0.1564 \\
        (H)w/o $L_{\mathrm{UserSSL2}}$ &0.3772 & 0.2514 & 0.4115 & 0.2626 & 0.2845 & 0.1565 \\
        (I)w/o $\beta$ &0.3797 & 0.2525 & 0.4109 & 0.2597 & 0.2826 & 0.1527 \\
        \bottomrule
        
    \end{tabular}
\end{table}

\noindent\textbf{Effect of Dual-level Interest Learning.} We design two variants to observe the impact of dual-level interest learning in CI4GI on performance. Variant A removes group-level interests in CI4GI, i.e., only item-level interests of users and groups are used to compute similarity. Variant B removes item-level interests in CI4GI.

The experimental results on three datasets are listed in the top part of Table~\ref{tab:ablation}. Removing either interest leads to significant performance degradation, which illustrates the effectiveness of dual-level interest modeling in CI4GI. On Steam, the performance degradation of variant A is more significant, indicating that group-level interests play a more important role. In contrast, on Mafengwo and Weeplaces, the performance degradation of variant B is more significant, i.e., item-level interests play a more important role. This may be due to the fact that Weeplaces has a very low average number of users per group (fewer than 3), making it difficult to accurately characterize groups using only group-level interests, leading to a substantial performance drop.


\noindent\textbf{Effect of Interest Enhancement Strategy.}
We build three variants to observe the effect of interest enhancement strategy during users' item-level interest learning in CI4GI. Variant C deletes the interest enhancement strategy, i.e., directly uses the user representation obtained through the GAT as the item-level interest representation of the user. Variant D removes the item representation enhancement, 
replacing the item representation in user-item GAT learning with random initialization.
Variant E deletes the contextual enhancement.

The experimental results of these three variants on three datasets are listed in the middle part of Table~\ref{tab:ablation}. The performance of variants D and E is lower than that of CI4GI, indicating that removing any of the interest enhancement methods results in a decrease in performance. Among the two variants, variant E shows the most significant performance degradation, highlighting the effectiveness of contextual enhancement in improving group identification. 
This is likely because it identifies the item that motivates the user to join the group as a potential interest, effectively enriching the user's item-level preferences.
Additionally, variant C exhibits a significant performance degradation, which suggests that the simultaneous deletion of the two interest enhancement methods has a significant negative impact on model performance.

\noindent\textbf{Effect of Contrastive Learning.}
We build four variants to observe the effect of contrastive learning in CI4GI. Variant F deletes the entire contrastive learning(CL) module. Variant G removes the vanilla contrastive learning loss $L_{\mathrm{UserSSL1}}$, variant H removes the contrastive learning loss with dynamic false-negative sample optimization $L_{\mathrm{UserSSL2}}$, and variant I deletes the annealing parameter $\beta$ used to balance the two contrastive learning losses, setting $\beta$ to a constant value of 0.5.

The experimental results on three datasets are listed in the bottom part of Table~\ref{tab:ablation}. 
Both variants G and H underperform compared to CI4GI, indicating that removing either contrastive learning loss negatively impacts model performance. Variant I experiences a more significant performance degradation compare to variants G and H, as the absence of the annealing parameter prevents the model from properly balancing the two contrastive learning losses. This causes the model to rely on the Wasserstein distance for identifying false-negative samples before it has sufficiently learned user interest patterns and distributional representations. As a result, the identified false-negative samples are heavily influenced by random initialization and lack meaningful guidance. Additionally, variant F shows a substantial performance decline, highlighting the crucial role of the contrastive learning module in CI4GI.

\subsection{Hyperparameter Sensitivity Analysis}

\begin{figure}[t]
    \centering
    \includegraphics[width=\linewidth]{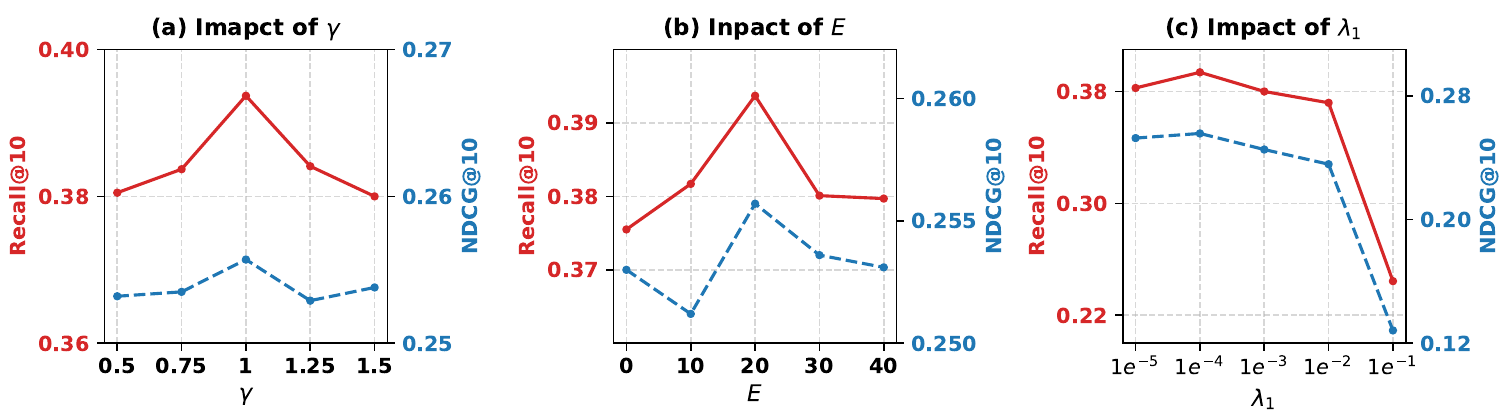}
    \caption{Sensitivity analysis of hyperparameters $\gamma$, $E$ and $\lambda_1$ on Mafengwo dataset.}
    \label{fig:hyper}
\end{figure}

We perform experiments on Mafengwo datasets to explore the sensitivity of the weight of contextual enhancement $\gamma$, the thershold $E$ in annealing parameters $\beta$ and the wight of contrastive learning loss $\lambda_1$. We fix other hyper-parameters, and tune $\gamma$, $E$ and $\lambda_1$ within \{0.5, 0.75, 1, 1.25, 1.5\}, \{0, 10, 20, 30, 40\}, \{$1e^{-5}$, $1e^{-4}$, $1e^{-3}$, $1e^{-2}$, $1e^{-1}$\}, respectively.
From the results in Fig. \ref{fig:hyper}, we can observe that as $\gamma$ increases from 0.5 to 1, the performance of CI4GI shows an increasing trend, and when $\gamma$ exceeds 1, the model performance starts to decrease. This suggests that appropriate contextual enhancement representation weight helps to improve the quality of the user representation, which in turn improves the model performance. 
The performance of CI4GI first rises and then falls as $E$ increases. 
When $E$ is too small, the model is prematurely subjected to unguided false negative sample identification before it has learned the basic user interests. Conversely, when $E$ is too large, the weight of contrastive learning with negative sample optimization is consistently small and fail to guide model optimization.
CI4GI achieves optimal performance when $\lambda_1=1\times e^{-4}$. If $\lambda_1$ is too small, the contrastive learning loss has little impact, while if $\lambda_1$ is too large, the weight of the contrastive learning loss is almost equal to the weight of the main loss, leading to a decrease in effectiveness.
\section{Conclusion}
\label{Conclusion}
Group identification is a challenging recommendation task, as a user's decision on joining a group is jointly influenced by both group-level and item-level interests. Therefore, effectively capturing these two types of interests and their collaborative relationship is crucial for accurate group identification.
We propose a model CI4GI, which simultaneously models users' group-level and item-level interests, and designs an interest enhancement strategy to capture the mutual enhancement of the dual-level interests through item representation enhancement and contextual enhancement. Meanwhile, we design a contrastive learning strategy with dynamic false-negative sample optimization to improve the alignment of cross-level interests. Experimental results on public datasets show that CI4GI significantly improves the accuracy of group identification.

\section*{Acknowledgment}
This work was supported by the National Natural Science Foundation of China under Grant No. 62072450. 
\end{document}